\documentstyle[editedvolume,numreferences,epsf]{crckapb}
\begin{opening}

\title{Charge Fluctuations and Dephasing in Coulomb Coupled Conductors} 

\author{Markus  B\"uttiker}

\institute{D\'epartement de Physique Th\'eorique, Universit\'e de Gen\`eve,\\
CH-1211 Gen\`eve 4, Switzerland}

\end{opening}
\runningtitle{Charge Fluctuations .....}
\begin{document}
\noindent
It is shown that the dephasing rate in Coulomb coupled mesoscopic structures
is determined by charge relaxation resistances. The charge relaxation 
resistance together with the capacitance determines the RC-time of 
the mesoscopic structure and at small frequencies determines the voltage 
fluctuation spectrum. Self-consistent expressions are presented which give 
the charge relaxation resistance and consequently the dephasing rate in terms 
of the diagonal and off-diagonal elements of a generalized Wigner-Smith 
delay-time matrix. Dephasing rates are discussed both for the equilibrium 
state and in the transport state in which charge fluctuations are 
generated by shot noise. A number of different geometries are discussed.  
This article is to appear in 
{\it Quantum Mesoscopic Phenomena and Mesoscopic Devices}, 
edited by I. O. Kulik and R. Ellialtioglu, (Kluwer, unpublished). 

\newpage

\section{Introduction} 
It is the purpose of this work to provide a 
self-consistent discussion of charge 
and potential fluctuations in Coulomb coupled mesoscopic conductors
and to apply the results to evaluate dephasing rates 
of Coulomb coupled conductors.  
Charge and potential fluctuation spectra
are important in a number of problems: the theory of dynamical 
(frequency-dependent) conductance of 
mesoscopic systems can be developed from a fluctuation theory \cite{btp}
(the dynamical conductance is then obtained from the 
fluctuation dissipation theorem and the Kramers-Kronig 
relations); already in the white noise limit, 
shot noise outside the ohmic range, 
is renormalized by charge fluctuations \cite{bb};
furthermore, the currents induced into gates capacitively 
coupled to a conductor, or a tunneling microscope tip (used as a small
movable gate) depend on the charge fluctuations of the mesoscopic 
conductor \cite{plb}. At low temperatures, the 
dephasing rates, which give the time over which 
a quasi-particle retains phase memory, are 
determined by potential fluctuations. 
It is clearly desirable to have a theory of charge fluctuations 
which works in all these situations.

The theory of fluctuations in mesoscopic conductors, 
has been predominantly concerned 
with current fluctuations \cite{kulik,khlus,leso,mb92}, and 
correlations of currents at different terminals of a multiprobe 
conductor \cite{mb92,henny,oliv}. 
For an extended review of shot noise 
in mesoscopic conductors, we refer the reader to Ref. \cite{physr}.
As the above mentioned rare examples \cite{btp,bb,plb} 
demonstrate, the theory can, however, 
be extended to treat charge and potential fluctuations.
In contrast to fluctuations in the total current at a contact 
of a mesoscopic sample, a discussion of charge fluctuations is more 
difficult, since it necessitates a treatment of interactions.

\begin{figure}
\vspace*{0.5cm}
\epsfysize=9cm
\epsfxsize=7cm
\centerline{\epsffile{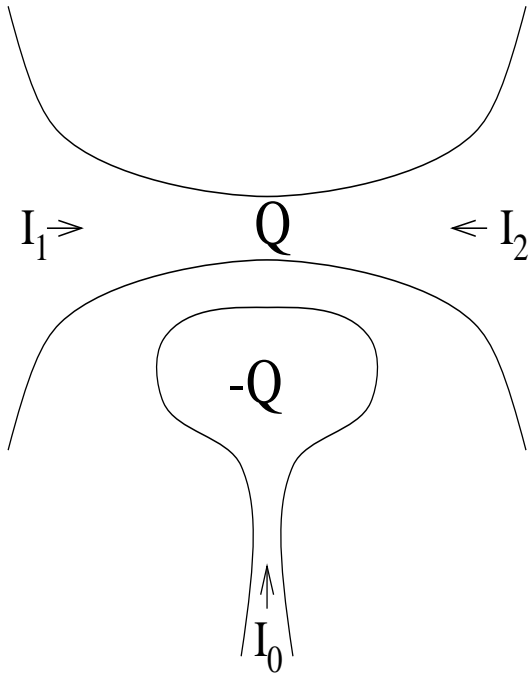}}
\vspace*{0.5cm}
\caption{ \label{ped}
Cavity with a charge deficit $-Q$ in the proximity of a quantum point contact 
with an excess charge $Q$. The dipolar nature of the charge distribution 
ensures the conservation of currents $I_{0}$, $I_{1}$ and $I_{2}$ 
flowing into this structure. After \protect\cite{plb} . 
}
\end{figure}
 
To emphasize the need for a self-consistent treatment 
of charge fluctuations, consider the conductor shown in Fig. \ref{ped}.
A quantum point contact \cite{vanwees,wharam} is capacitively coupled to a second
mesoscopic conductor which might represent a gate or 
a small quantum chaotic cavity coupled only via a single lead 
to an electron reservoir \cite{plb}.  
The currents to the right and left 
of the quantum point contact (QPC) are $I_{1}$ and $I_{2}$
and the current that flows from the cavity to its electron reservoir 
is denoted by $I_{0}$. 
We want to assume that the QPC and the cavity are so close to one another
that every electrical field line which emanates from the cavity 
ends either again on the cavity or on the QPC. 
There exists then a volume which encloses 
these two conductors with the property that the net electric flux 
through the surface of this volume is zero. According to Gauss, the 
net electric charge inside this volume is thus zero. Hence any 
charge fluctuation $Q_{1}$ on the cavity must 
be counter-balanced by a charge fluctuation $Q_{2}$ on the QPC  
such that the total charge is preserved, $Q_{1} +Q_{2} = 0$.
Hence the charge on the cavity $Q_{1} \equiv - Q$
is totally correlated with the excess charge $Q_{2} \equiv Q $
on the QPC. 
The excess charge is dipolar with a charge accumulation 
on one of the conductors and a charge depletion on the other conductor. 
It is the conservation (and complete correlation) of the excess charges
on the two conductors which ensures the conservation of currents, 
\begin{equation}
I_{0} (t) + I_{1}(t)  + I_{2}(t) = 0 .
\label{f0}
\end{equation} 
This conservation holds at any instant of time. 
A fluctuation theory based on independent particles, cannot 
describe the correlations in the charge fluctuations 
which are necessary to establish Eq. (\ref{f0}). 
In an independent particle approach the number of particles 
in the cavity would be independent of the number of
particles on the QPC. Thus an approach which takes the Coulomb 
interactions into account is necessary to 
establish such a basic property like current conservation.

Dephasing rates 
in disordered conductors are mostly discussed in connection with 
weak localization \cite{aak,sivan,blant}. 
Weak localization is a quantum effect which  
arises from the interference of 
time-reversed particle trajectories and which survives 
ensemble averaging.
The calculations of the dephasing rate for 
this effect is performed by first ensemble averaging 
such that the dynamics is effectively 
diffusive. 
This permits 
a treatment of the Coulomb interactions for a sample
that is uniform on the scale of the 
elastic scattering length: screening is treated with the help 
of a frequency and wave-vector dependent 
dielectric constant. The charge-fluctuations are essentially 
electron-hole pairs which leave the total charge on the conductor 
invariant. 

In contrast, the dephasing rates discussed here, {\it cannot} be 
applied 
to weak localization.  
The main effect which is investigated comes from 
carriers which leave or enter the conductor.
These carriers thus change the total charge 
on the conductor. It is clear that for the rates considered here 
the connections of the sample to the outside world, the properties 
of its contacts, play an important role. This is again in contrast 
to dephasing rates determined by electron-hole pair
excitations. Both approaches have in common that the charge excitations 
considered are dipolar: here we place the electron on one conductor
and the hole on the other conductor. 

The experiments we have in mind are provided in
recent works by Buks et al. \cite{buks1} and by Sprinzak et al. \cite{buks2}. 
In these 
experiments, the effect of a current carrying QPC  on a Coulomb coupled 
nearby phase-coherent system is investigated. The dephasing rate which 
is measured is proportional to the voltage applied to the QPC. 
In addition to the discussion provided in the experimental works, 
dephasing in Coulomb coupled structures, has found 
attention in a number of theoretical works \cite{gurvitz,aleiner1,levinson}. 
Widely different approaches
have been used, but the point of view provided here, which 
emphasizes the correlations of the charge fluctuations on the two 
conductors and the need for a self-consistent discussion, seems novel. 
A brief discussion of the results of our work is presented in Ref. \cite{mbam}
for the geometry of the experiment of Sprinzak et al. \cite{buks2}. 
Different and closely related aspects of the work presented here 
are discussed in two conference proceedings \cite{mori,sitges}. 
The experiment \cite{buks2} is also discussed in Ref. \cite{levins2} but 
without an attempt to provide a self-consistent 
discussion of charge fluctuations.

\section{The mesoscopic capacitor (macroscopic backgate)}

\begin{figure}
\epsfysize=9cm
\epsfxsize=7cm
\centerline{\epsffile{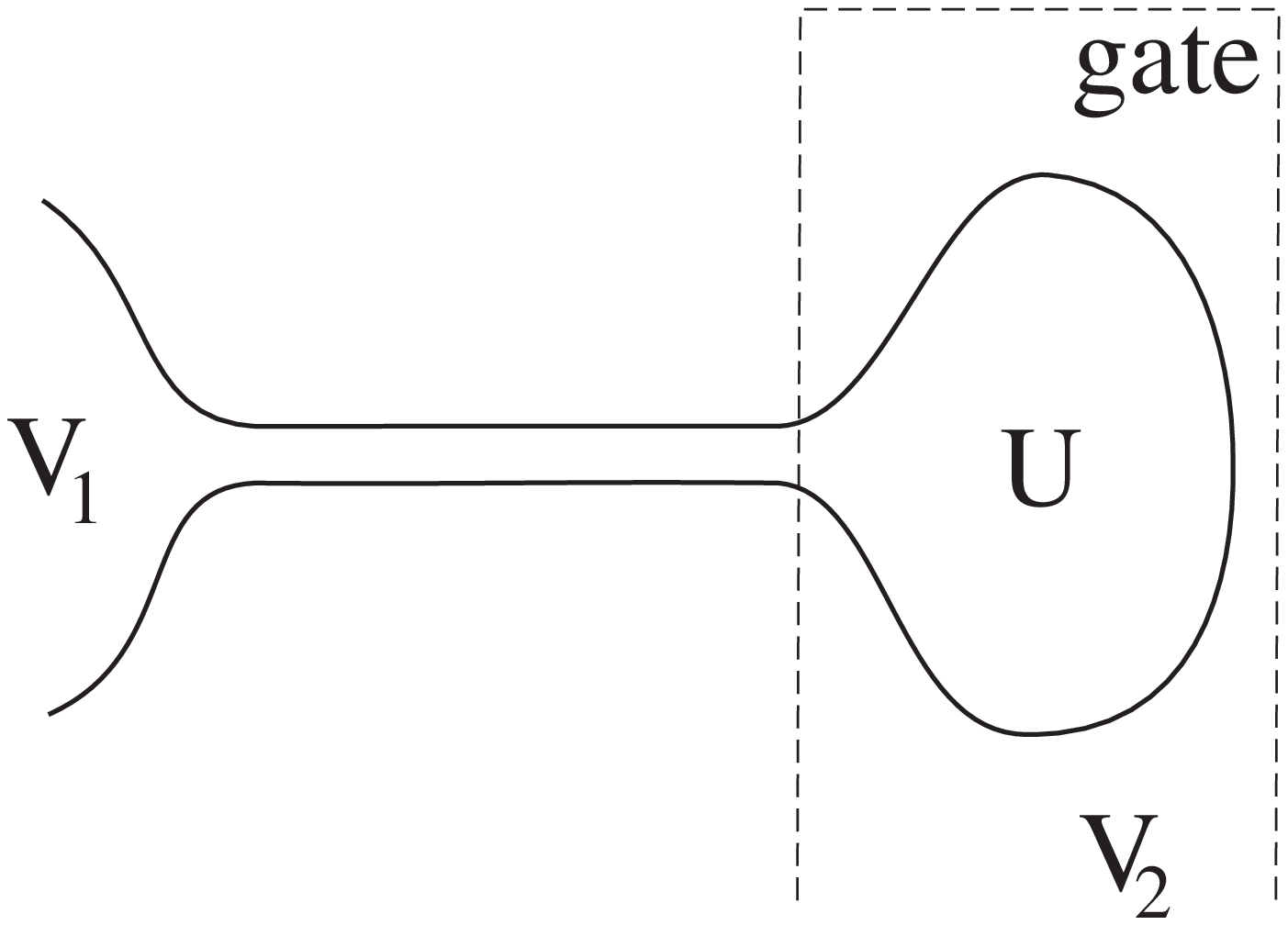}}
\vspace*{0.3cm}
\caption{ \label{mesocap}
Mesoscopic capacitor connected via a single lead to an electron reservoir 
and capacitively coupled to a gate. $V_1$ and $V_2$ are the potentials applied
to the contacts, $U$ is the electrostatic potential of the cavity. 
After \protect\cite{korea}. }
\end{figure}

The simplest system for which it is instructive to investigate the 
fluctuations of charge is a mesoscopic capacitor \cite{btp}. 
Fig. \ref{mesocap} shows a cavity capacitively coupled to a backgate and connected via 
a single lead to an electron reservoir. In a first
step we will 
treat the gate as a macroscopic conductor. 
We assume that the electrostatic potential on the cavity can be described 
be a single parameter $U$. The theory presented below 
is not limited to this simplifying assumption but can be extended 
to treat the microscopic landscape \cite{math}. 
In the presence 
of an oscillating potential 
at the reservoir with Fourier amplitude $V_{1}(\omega)$ 
or a potential $V_{2}(\omega)$ at the gate, the electrostatic
potential on the cavity oscillates with an amplitude $U(\omega)$. 
These potential oscillations are connected to the charge oscillations 
$Q(\omega)$ on the cavity via $Q(\omega) = C (U(\omega) - V_{2}(\omega))$. 
Here $C$ is the geometrical capacitance coupling the charge on the cavity 
to that on the gate. 
The resulting dynamic conductance, 
that relates the ac-current through this 
structure to a small ac-voltage applied between the reservoir 
and the backgate is \cite{btp}  
\begin{equation}
G(\omega) =  \frac{-i \omega C_{\mu}}
{1 - i \omega R_{q}C_{\mu}} .
\label{rcg}
\end{equation}
In Eq. (\ref{rcg}) we have retained only the pole in the complex
frequency-plane with the smallest 
imaginary part. 
The dynamic conductance of the mesoscopic capacitor is, like that 
of a macroscopic capacitor, determined by an $RC$-time. 
But instead of only purely classical quantities, we obtain now 
expressions that contain quantum corrections due to the phase-coherent
electron motion in the cavity. It turns out 
that the $RC$-time can be expressed with the help 
of the Wigner-Smith time-delay matrix \cite{smith}  
\begin{equation}
{\bf \cal{N}} = \frac{1}{2\pi i} {\bf s}^{\dagger} \frac{d{\bf s}}{dE} . 
\label{ws}
\end{equation}
Here ${\bf s}$ is the scattering matrix which relates 
the incident current amplitudes in the lead connecting the cavity
to the reservoir $1$ to the out-going current amplitudes 
in this lead (see also Appendix A). 
The sum of the diagonal elements of this  matrix 
determines the density of states \cite{dashe}
\begin{equation}
N = Tr{\cal N} = \frac{1}{2\pi i} Tr[{\bf s}^{\dagger} \frac{d{\bf s}}{dE}]
\label{den}
\end{equation}
and gives rise to a "quantum capacitance" $e^{2} N$
which in series with the geometrical capacitance 
determines the electrochemical capacitance \cite{btp}
\begin{equation}
C_{\mu}^{-1} = C^{-1} + (e^{2} N)^{-1} . 
\label{cmu}
\end{equation}
The resistance which counts is the charge relaxation 
resistance \cite{btp} 
\begin{equation}
R_{q} = \frac{h}{2e^{2}}\frac{Tr[{\cal N}^{\dagger} {\cal N}]}
{[Tr{\cal N}]^{2}} .
\label{rq}
\end{equation} 
For simplicity we have given these results, Eqs. (\ref{ws} - \ref{rq}),
only in the zero temperature limit. 
It is instructive to consider a basis in which the 
scattering matrix is diagonal. 
Since we have only reflection, all eigenvalues of the scattering matrix 
are of the form $exp(i\zeta_{n})$ where 
$\zeta_{n}$ is the phase which a carrier 
accumulates from the entrance to the cavity through multiple 
scattering inside the cavity until it finally exits the cavity.
Thus the density of states can also be expressed as 
\begin{equation}
N = (1/2\pi) \sum_{n}(d\zeta_{n}/dE)
\label{denp}
\end{equation} 
and is seen to be proportional to the total Wigner time delay carriers 
experience in the cavity. The time delay for channel $n$ is \cite{note1}
$\tau_n = \hbar d\zeta_n/dE$. 
Similarly we can express the charge relaxation resistance in terms 
of the energy derivatives of phases
and we obtain in the zero-temperature limit, 
\begin{equation}
R_{q} = \frac{h}{2e^{2}}\frac{\sum_{n} (d\zeta_n/dE)^{2}}{[\sum_n d\zeta_{n}/dE]^{2}}.
\label{rqp}
\end{equation} 
$R_q$ is thus determined by the sum of the squares of the delay times
divided by the square of the sum of the delay times. 

We now briefly discuss these results. 
First, our Eq. (\ref{cmu}) for the electrochemical capacitance predicts 
that it is not a purely geometrical quantity but that 
it depends on the density of states of the cavity. 
This effect is well known from investigations of the capacitance of 
the quantized Hall effect. More recent work investigates 
the mesoscopic capacitance of quantum dots and wires
and is often termed {\em capacitance spectroscopy}. In addition
to describing the average behavior, our results can also be used 
to investigate the fluctuations in the capacitance. 
Similar to the universal conductance fluctuations 
there are capacitance fluctuations in mesoscopic samples
due to the fluctuation of the density of states. 
Such effects can be expected to be most pronounced 
if the contact permits just the transmission of a single channel. 
Then it is necessary not only to investigate the fluctuations 
of the mean square fluctuations but the entire distribution function. 
Such an investigation was carried out by Gopar et al. \cite{gopar}
in the single channel limit and by Brouwer and the author \cite{brbu}, 
and Brouwer et al. \cite{bfb} for chaotic cavities with quantum point contacts 
which are wide open (many channel limit). 
Since in experiments \cite{marcus}
the Coulomb energy $e^{2}/C$ is typically much larger than 
the level separation $\Delta$ these fluctuations are small.
A very interesting prediction 
which follows from the density 
of states dependence of the electrochemical capacitance is that for a sample 
of the form of a loop with an Aharonov-Bohm flux through the hole 
of the loop, the capacitance should exhibit Aharonov-Bohm 
oscillations \cite{physica}. Aharonov-Bohm oscillations in the 
capacitance of small rings have recently been measured by 
Deblock et al. \cite{deblock}.  

Next let us discuss briefly the charge relaxation resistance $R_q$. 
An over-
view of charge relaxation resistances in mesoscopic systems 
is presented in Ref. \cite{korea}. 
First we note that the resistance unit is not the resistance quantum 
$h/e^{2}$ but $h/2e^{2}$. The factor two arises since the cavity 
is coupled to one reservoir only. Thus only half the energy is dissipated 
as compared to dc-transport through a two terminal conductor. 
Second, we note that in the single channel limit, Eq. (\ref{rqp}) 
is {\em universal} and given just by $h/2e^{2}$. This is astonishing since 
if we imagine that a barrier is inserted into the lead connecting 
the cavity to the reservoir one would expect a charge relaxation resistance 
that increases as the transparency of the barrier is lowered. 
Indeed, if there is a barrier with transmission probability $\cal T$ 
per channel in the lead connecting the cavity and the reservoir, 
then in the large channel limit, for ${\cal T}M \gg 1$, $R_q$ is \cite{bunp}, 
\begin{equation}
R_q = (h/e^{2}) (1/{\cal T}M).
\label{bee} 
\end{equation}  
In the large channel-number limit, Eq. (\ref{rqp})
is proportional to $1/M$, where $M$ is 
the number of scattering channels, 
and proportional to $1/{\cal T}$.  
Thus in the large 
channel limit Eq. (\ref{rqp}) behaves as expected. 

Let us next consider the fluctuation spectra of the current, charge 
and potential. Ref. \cite{btp} gives a derivation of these 
spectra using a dynamic fluctuation theory of mesoscopic conductors. 
The current fluctuation spectra must, however, in any case, 
be connected to the dynamical conductance via the fluctuation-dissipation 
theorem. This gives for the current noise spectrum \cite{btp}, 
\begin{equation} \label{sii}
S_{II}(\omega) =  2kT \frac{\omega^{2} C_{\mu}^{2} R_{q}}
{1 +\omega^{2} R^{2}_{q}C^{2}_{\mu}} .
\end{equation}
Since the charge on the cavity is the time-derivative of the current, 
we find for the fluctuation spectrum of the total charge on the cavity, 
\begin{equation}
S_{QQ}(\omega) =  2kT \frac{C_{\mu}^{2} R_{q}}
{1 +\omega^{2} R^{2}_{q}C^{2}_{\mu}} ,
\label{sqq}
\end{equation}
and since the charge is related via the geometrical capacitance 
to the potential, $Q(\omega) = C U(\omega)$ we find 
\begin{equation}
S_{UU}(\omega) =  2kT \frac{C_{\mu}^{2}}{C^{2}} \frac{R_{q}}
{1 +\omega^{2} R^{2}_{q}C^{2}_{\mu}} .
\label{suu}
\end{equation} 
Let us pause here and consider two limiting cases. 
First, if the charging energy is unimportant, the geometrical capacitance 
becomes very large, and the electrochemical capacitance is essentially
determined by the quantum capacitance $C_{\mu} \simeq e^{2}N$.
In this case the spectrum for the voltage fluctuations, Eq. (\ref{suu}), tends to zero. 
There is no screening of the charge pile-up. In the opposite limit, 
when the geometrical capacitance tends to zero, the quantum capacitance 
becomes unimportant and $C_{\mu} \simeq C$. In this case, both 
the current fluctuation spectrum and the charge fluctuation spectrum 
become very small. Charging of the cavity is now energetically very 
expensive, and we can neither drive a current through the structure 
nor can we pile-up charge.  In this limit,
the spectrum of the potential fluctuations as a function of the geometrical 
capacitance reaches its maximum amplitude, whereas its width in frequency 
becomes increasingly narrow. 

The reason that in the small capacitance (charge neutral limit) 
both the current and the charge fluctuation spectrum tend to zero, 
whereas the voltage fluctuation spectrum stays finite, is also due 
to electron-hole pair excitations. If simultaneously an electron and 
hole enter the cavity, the total charge remains unchanged, there is no
net current generated but there are nevertheless potential fluctuations. 

Let us next consider the dephasing of a carrier in the mesoscopic 
cavity due to the potential fluctuations. We follow Ref. \cite{levinson} 
and relate the 
phase $\phi$ of a carrier to the fluctuations in the potential 
via, $i\hbar d\phi/dt = eU(t)$. Integrating this equation, 
defines for $t \gg RC$ a dephasing rate 
\begin{equation}
\Gamma_{\phi} = \frac{\langle (\phi(t) - \phi(0))^{2} \rangle}{t} = 
\frac{1}{t} \frac{e^{2}}{\hbar^{2}} 
\langle \int_{0}^{t} dt^{\prime} (U(t^{\prime}) - U(0))^{2} \rangle
\label{dr1}
\end{equation} 
which is related to the zero-frequency limit of the 
the voltage fluctuation spectrum via, 
\begin{equation}
\Gamma_{\phi} = 
(e^{2}/2\hbar^{2}) S_{UU}(0) .
\label{gammaphi}
\end{equation} 
Thus up to fundamental constants, the dephasing rate is determined 
by the white noise limit of the potential fluctuation spectrum. 
Using Eq. (\ref{suu}) 
for the mesoscopic capacitor considered, we thus obtain, 
\begin{equation}
\Gamma_{\phi} = 
(e^{2}/\hbar^{2}) kT ({C_{\mu}^{2}}/{C^{2}}) R_{q}. 
\label{mctau}
\end{equation} 
This result expresses the dephasing rate in terms of {\it electrical} 
quantities. 
The dephasing rate is essentially determined by the charge 
relaxation resistance $R_{q}$ and a ratio of capacitances. 
We have already noticed, that in the free-electron limit, 
the voltage fluctuation spectrum vanishes, since $C_{\mu}/C$ tends to zero.
Consequently, in this limit there is no phase-breaking. 
In the charge neutral limit, the dephasing rate is determined 
by the charge relaxation resistance alone, since 
${C_{\mu}^{2}}/{C^{2}}$ tends to $1$. It is this later limit, that 
describes most often the physical situation which we encounter 
in mesoscopic conductors. The charging energy $E_c = e^{2}/2C$ is typically 
an order of magnitude larger than the level spacing \cite{marcus}. 
Thus the dephasing rate is essentially determined by the charge 
relaxation resistance. Consequently, 
the knowledge we have gained on charge relaxation resistances can 
immediately be applied to the dephasing time. 

We re-emphasize that the dephasing rate, 
as given by Eqs. (\ref{dr1}, \ref{mctau}),
cannot be compared with a weak localization dephasing time. 
We considered here only a uniform potential inside the cavity, 
and a uniform potential does not affect a weak localization 
loop, since both the backward and forward trajectories experience 
the same potential. Eqs. (\ref{dr1}, \ref{mctau}) are sample 
specific results and as has been pointed out before are due to carriers 
entering or leaving the sample. 

We must leave it as an open question, to what extent
the dephasing time introduced above is physically 
meaningful. For small conductors, 
$R_q$ is not well defined by its average nor its mean square fluctuations, 
but must be characterized by an entire distribution. 
Thus for small 
samples, it is similarly only meaningful to 
define a distribution of dephasing rates. 
But if the dephasing rate can only be characterized in terms 
of a distribution function, it becomes a less useful object. 
In principle, rather than defining a rate, it is 
desirable to evaluate the quantity of interest, 
such as the density of states of the cavity, the conductance 
which we consider next, directly, without relying 
on the concept of a dephasing rate. 

\section{Role of external impedance} 

The fluctuations in a mesoscopic conductor 
depend not only on the conductor itself, but also on the impedance of the 
external circuit. The considerations given above apply only for the case 
of vanishing external impedance. Let us now briefly discuss the case 
where the external impedance $Z_{ext}(\omega)$
does not vanish. The entire circuit can be described by 
writing two Langevin equations \cite{btp,mb92} for the mesoscopic conductor 
and for the external circuit. 
For the mesoscopic conductor we have 
\begin{equation}
\Delta I(\omega) = G(\omega) V(\omega) + \delta I(\omega) .
\label{l1}
\end{equation}
Here $G(\omega)$ is determined by Eq. (\ref{rcg}),
$V(\omega)$ is the voltage drop between reservoir and gate,  
and $\delta I(\omega)$ are the current fluctuations 
for infinite impe-
dance with a spectrum determined 
by Eq. (\ref{sii}). The fluctuating current in the external circuit 
is 
\begin{equation}
\Delta I(\omega) = - Z^{-1}_{ext}(\omega) V(\omega) + \delta I_{ext}(\omega) ,
\label{l2}
\end{equation}
where $\delta I_{ext}(\omega)$
has a spectrum 
$S_{ext} (\omega) = 2kT/Z_{ext}(\omega)$. 
Consider the important case where the external impedance 
is just a dc-resistance $R_{ext}$. 
The resulting current-, charge-, and voltage-fluctuation 
spectrum can be found by replacing in Eqs. (\ref{sii}-\ref{suu}),
the charge relaxation resistance $R_q$ by $R_q + R_{ext}$.
An external resistive impedance leads to an increase in the 
RC-time. As a consequence the dephasing rate is given by 
\begin{equation}
\Gamma_{\phi} = 
(e^{2}/\hbar^{2}) kT ({C_{\mu}^{2}}/{C^{2}}) (R_{q} + R_{ext}). 
\label{gaext}
\end{equation} 
A finite external resistive impedance increases the voltage fluctuations 
and shortens the dephasing time. The impedance conditions of the external
circuit are thus important if we want to make a quantitative 
comparison with experiment. Below, we now return to investigate
charge fluctuations and dephasing rates for the case of 
zero-impedance external circuits.

\section{Equilibrium dephasing in multilead systems}

\begin{figure}
\epsfysize=9cm
\epsfxsize=7cm
\centerline{\epsffile{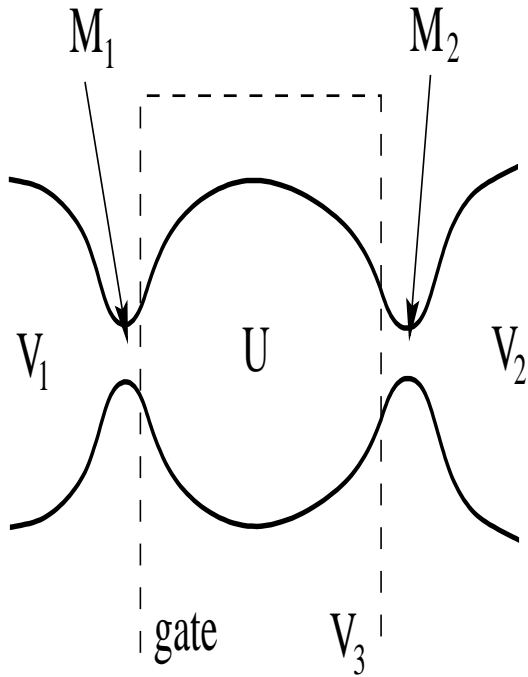}}
\vspace*{0.3cm}
\caption{ \label{twoprobe}
Two probe conductor connected to two electron-reservoirs and 
capacitively coupled to a gate. After Ref. \protect\cite{korea}. }
\end{figure}

Consider next a mesoscopic conductor 
that is connected to two or more reservoirs 
and capacitively coupled to a backgate. As an example,  
Fig. \ref{twoprobe} shows a chaotic cavity which is 
connected via two point contacts 
with $M_{1}$ and $M_{2}$ 
open channels to two reservoirs and capacitively coupled to a back gate. 
The equilibrium charge fluctuations in this conductor 
correspond to a situation in which all the potentials at all 
contacts are at equilibrium. Thus for the determination 
of the equilibrium charge fluctuations (and for a zero
impedance external circuit), it is irrelevant whether 
we consider the leads connected to different reservoirs 
or consider all the leads of the conductor connected to the same reservoir. 
But if all the leads are connected to the same reservoir it is evident 
that the theory presented above for the case of a single 
lead also applies to a conductor with many leads.
All that is necessary is to use the full scattering matrix in the 
evaluation of the density of states and of the charge 
relaxation resistance.  
Therefore, for the discussion of equilibrium charge relaxation resistances
we only need to distinguish the dimension of the scattering matrix 
(number of quantum channels) but not the number of leads. 

Following is a summary of what is known on charge 
relaxation resistances for various mesoscopic systems 
(see also Ref. \cite{korea} for an overview): 

For a capacitor with a single channel lead the charge relaxation 
resistance is universal and given by $R_q = h/2e^{2}$. 
For a nearly charge neutral cavity such that $C_{\mu} = C$, 
Eq. (\ref{mctau}) gives  
\begin{equation}
\Gamma_{\phi} = 
(4\pi^{2}/h) kT . 
\label{gaext1}
\end{equation} 
For a chaotic cavity 
connected via two {\it perfect single channel} leads to 
a reservoir the charge relaxation resistance $R_q$ 
must be characterized by a distribution (which based on 
Dyson's circular ensemble can be given analytically \cite{plb}). 
Consequently, the dephasing rate is also given by a distribution 
function $P(\Gamma_{\phi}/\Gamma_{0})$. 
Here $\Gamma_{0}= (4\pi^{2}/h) kT$. For the orthogonal ensemble 
the symmetry parameter is $\beta = 1$ and for the unitary ensemble it is 
$\beta = 2$. For these two symmetry classes, Ref. \cite{plb} finds
a distribution function which is non-zero only between 
$\Gamma_{\phi}/\Gamma_0 = 1/4$ and $\Gamma_{\phi}/\Gamma_0 = 1/2$
and is given by 
\begin{equation}
    P(\Gamma_{\phi}/\Gamma_{0}) = \left\{ 
	\begin{array}{ll}
	    4, & \beta=1, \\
	    30(1-2\Gamma_{\phi}/\Gamma_{0})\sqrt{4\Gamma_{\phi}/\Gamma_{0}-1}, 
	    & \beta=2.
	\end{array}
	\right.
	\label{distr}
\end{equation}
For a chaotic cavity (see Fig. \ref{twoprobe})
coupled to leads with many channels $M = M_{1} +M_{2}$,
$R_q$ becomes well defined and exhibits only small fluctuations 
around its average \cite{brbu}. Here $N$ is the total number of channels of all contacts 
and $M_{1}$ and $M_{2}$ are the number of channels for a chaotic 
cavity coupled by contacts with $M_{1}$ and $M_{2}$ channels to reservoirs. 
On the ensemble average $R_q = (h/e^{2}) (M_{1} + M_{2})^{-1}$. 
Note that the conductances of the two quantum point contacts 
add in {\it parallel}. This is in contrast to the (ensemble
averaged) conductance of the cavity 
which is $ G^{-1} = (h/e^{2})
(M_{1}^{-1} + M_{2}^{-1})$ and corresponds to the series addition 
of the resistances of the contacts. Thus whether or not 
the effect discussed here contributes to the 
dephasing rate in a chaotic cavity is easy to test experimentally: 
Whereas for two very unequal contacts $M_{1} \ll M_{2}$ 
the resistance is dominated by the smaller contact, (i. e. it is 
determined by $M_{1}$), the charge relaxation resistance and thus 
the dephasing rate discussed here is dominated by 
the larger contact $M_{2}$. 

For a perfect ballistic one-channel wire \cite{bhb} connecting two reservoirs, 
and capacitively coupled to a gate, the charge 
relaxation resistance is $R_q = h/4e^{2}$. For a two dimensional
wire subject to a high magnetic field, such that the only extended states
at the Fermi energy are edge states, and for a geometry in which 
a gate couples  
to one edge state \cite{chen}, 
the charge relaxation resistance is \cite{chris}
$R_{q} = h/2e^{2}$. 
If more than one edge state contributes to transport,
$R_q$ depends on the density of states of the different edge channels.  
Of much interest is the charge relaxation resistance of a quantum point contact 
and we discuss it now in some detail.

\section{Charge relaxation resistance of a quantum point contact}

We consider a quantum point contact (QPC) formed with the 
help of gates such that it connects two 
two-dimensional electron gases \cite{vanwees,wharam}.  
Here we treat these gates as macroscopic.  
For simplicity, we consider a
symmetric contact: We assume that the electrostatic potential is 
symmetric for electrons approaching the contact from the left or from 
the right. This implies that the two gates 
are located symmetrically. We can combine the capacitances 
of the conduction channel to the two gates and consider 
a single gate as schematically shown in  Fig.~\ref{ped}.
For a symmetric scattering potential  
the scattering matrix (in a basis in which the 
transmission and reflection matrices are diagonal) is 
for the $n$-th channel of the form
\begin{equation}
    s_{n}(E) = \left( \begin{array}{ll}
	-i \sqrt{R_{n}} \exp(i\phi_{n}) & \sqrt{T_{n}} \exp(i\phi_{n}) \\
	\sqrt{T_{n}} \exp(i\phi_{n}) & -i\sqrt{R_{n}} \exp(i\phi_{n})
	\end{array} \right) ,
	\label{sqpc}
\end{equation} 
where $T_{n}$ and $R_{n}= 1-T_{n}$
are the transmission and reflection probabilities 
and $\phi_{n}$ 
is the phase accumulated by a carrier
during reflection or transmission at the QPC. 
For the eigenvalues  $exp(i\zeta_{n\pm})$ of this matrix a little algebra 
shows that the derivatives with respect to energy of the eigenphases
$\zeta_{n\pm}$ 
are given by 
\begin{equation}
\frac{d\zeta_{n\pm}}{dE} = \frac{d\phi_{n}}{dE} 
\pm \frac{1}{2T_{n}^{1/2}R_{n}^{1/2}} \frac{dT_{n}}{dE} . 
\label{eigen} 
\end{equation}
For the density of states, Eq. (\ref{den}), this leads to 
$N = (1/2\pi) \sum_{n} (d\zeta_{n+} + d\zeta_{n-}) = (1/\pi) 
\sum_{n} d\phi_{n}/dE$, and for the charge relaxation resistance, 
Eq. (\ref{rq}), we find
\begin{equation}
R_{q} = \frac{h}{4e^{2}} 
\frac{\sum_{n} \left[(\frac{d\phi_{n}}{dE})^{2} + 
\frac{1}{4T_{n}R_{n}} (\frac{dT_{n}}{dE})^{2} \right]}
{[\sum_{n} \frac{d\phi_{n}}{dE}]^{2}} .
\label{rqgen}
\end{equation}

If only a few channels are open the potential 
has in the center of the conduction channel 
the form of a saddle \cite{mbqpc}:
\begin{equation}
    V(x,y) = V_0 + \frac{1}{2} m \omega_y^2 y^2
    - \frac{1}{2} m \omega_x^2 x^2 ,
\end{equation}
where $V_0$ is the electrostatic potential at the saddle and the curvatures of
the potential are parametrized by $\omega_x$ and $\omega_y$.
The resulting transmission probabilities
have the form of Fermi functions  
$T(E) \equiv f (E) = 1/(e^{\beta (E-\mu)}+ 1)$ 
(with a negative temperature 
$\beta = - 2 \pi /\hbar \omega_x$
and 
$\mu = \hbar \omega_y (n+1/2) +V_{0})$.
As a function of energy (gate voltage) the conductance rises 
step-like \cite{vanwees,wharam}. 
The energy derivative of 
the transmission probability 
\begin{equation}
dT_{n}/dE = - \beta f(1-f) =  (2 \pi /\hbar \omega_x) T_{n}(1-T_{n}) 
\end{equation}
is itself proportional to the transmission probability times 
the reflection probability. We note that such a relation 
holds not only for the saddle point model of a QPC but also for 
instance for the adiabatic model of Glazman et al. \cite{glaz}. 
As a consequence 
\begin{equation}
\frac{1}{4T_{n}R_{n}} (dT_{n}/dE)^{2} 
= (\frac{\pi}{\hbar \omega_x})^{2} T_{n}R_{n}
\label{dtde}
\end{equation} 
is proportional to $T_{n}R_{n}$. 
Thus we can re-write the charge relaxation resistance of a saddle 
QPC as 
\begin{equation}
R_{q} = \frac{h}{4e^{2}} 
\frac{\sum_{n} \left[(\frac{d\phi_{n}}{dE})^{2} +  
(\frac{\pi}{\hbar \omega_x} T_{n}R_{n})^{2}\right]}
{[\sum_{n} \frac{d\phi_{n}}{dE}]^{2}} . 
\label{rqsad}
\end{equation}

\begin{figure}
\epsfysize=8cm
\epsfxsize=7cm
\centerline{\epsffile{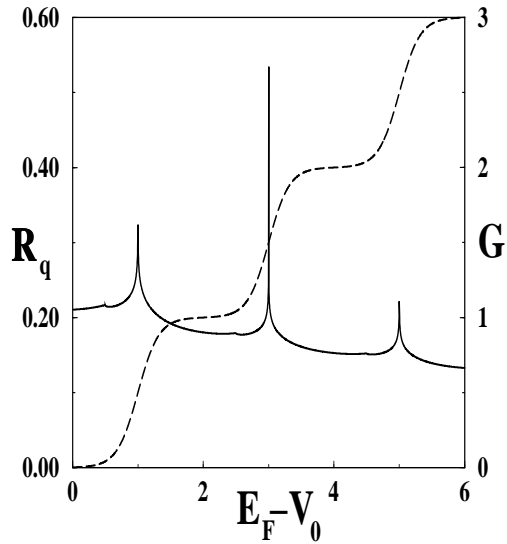}}
\caption{ \label{rqfig}
Charge relaxation 
resistance $R_{q}$ of a saddle QPC in units of $h/4e^{2}$ for 
$\omega_y/\omega_x = 2$ and a screeining length of 
$m\omega_x \lambda ^{2} /\hbar = 25$ as a function of $E_F -V_{0}$
in units of $\hbar \omega_x$ (full line). 
The broken line shows the conductance of the QPC. 
After Ref. \protect\cite{ammb} .    
}
\end{figure}

To obtain the density of states 
of the $n$-th eigenchannel, we use the relation between 
density and phase (action) and $N_{n} = (1/\pi) d\phi_{n}/dE$. 
We evaluate the phase semi-classically.
The spatial region of interest for which we have to find 
the density of states is the region over which the electron
density in the contact is not screened completely. 
We denote this length by $\lambda$. 
The density of states is then found from  
$N_{n}=1/h \int_{-\lambda}^{\lambda} \frac{dp_n}{dE} dx$ 
where $p_{n}$ is the classically allowed momentum.
A simple calculation gives a density of states \cite{plb}
\begin{equation}
N_{n} (E) = \frac{4}{h\omega_x}
asinh \left( \sqrt{\frac{1}{2} \frac{m\omega_x^2}{E-E_n}}\lambda  \right),
\end{equation}
for energies $E$ exceeding the channel threshold $E_n$ and gives 
a density of states 
\begin{equation}
N_{n} (E) = \frac{4}{h\omega_x}	    
acosh \left( \sqrt{\frac{1}{2} \frac{m\omega_x^2}{E_n-E}}\lambda \right), 
\end{equation}
for energies in the 
interval $E_n - (1/2) m \omega_x^2 \lambda \leq E < E_n$ 
below the channel threshold. 
Electrons with energies less than $E_n - \frac{1}{2} m \omega_x^2 \lambda $
are reflected before
reaching the region of interest, and thus do not contribute to the DOS.
The resulting density of states has a logarithmic singularity
at the threshold $E_{n}= \hbar\omega_y(n+\frac{1}{2})+V_0$  
of the n-th quantum channel. (A fully quantum mechanical 
calculation gives a density of states which exhibits also
a peak at the threshold but which is not singular). The charge relaxation 
resistance for a saddle QPC is shown in Fig. \ref{rqfig} for a set 
of parameters given in the figure caption. The charge relaxation 
resistance exhibits a sharp spike at each opening of a quantum channel. 
Physically this implies that the relaxation of charge, determined by 
the $RC$-time is very rapid at the opening of a quantum channel.

\section{Charge Fluctuations and the Scattering Matrix} 

We now present an approach which can be used to 
determine the fluctuations of the charge on Coulomb 
coupled conductors not only 
in the equilibrium state 
of the conductors but also 
if one or both conductors are in a transport state. 
First, we are concerned with the fluctuations 
of the total charge. Later on we will also derive expressions 
which permit the investigation of the local charge fluctuations. 

We have already pointed out in the preceding discussion, the fact 
that interaction effects are very important for the discussion of charge 
and potential fluctuations. For a moment, however, we now consider 
non-interacting carriers and only later on we introduce screening. 
First we want to derive an operator for the total charge on a mesoscopic 
conductor. To this end we use the continuity 
equation 
and integrate it over the total volume of the 
conductor. This gives a relation 
between the charge in this volume and the particle 
currents entering this volume, 
\begin{equation} \label{qtot}
\sum_{\alpha} \hat I_{\alpha}(\omega) =  
i\omega e \hat{\cal N} (\omega) , 
\end{equation}
where $\hat I_{\alpha}(\omega)$ 
is the current operator in lead $\alpha$ (see Appendix A) 
and $e \hat{\cal N}$ is the operator 
of the charge in the mesoscopic conductor. 
From the current operator (see Appendix A) 
we obtain \cite{plb} for the density operator 
\begin{equation} \label{qop}
\hat{\cal N}(\omega) = {\hbar} \sum_{\alpha\beta\gamma}
\sum_{mn} \int dE \hat a^{\dagger}_{\beta m}(E) 
{\cal N}_{\beta\gamma mn} (E, E + \hbar \omega) 
\hat a_{\gamma n} (E + \hbar \omega),
\end{equation}
whrere ${\hat a}^{\dagger}_{\beta m}(E)$ 
(and  ${\hat a}_{\beta m}(E)$) creates (annihilates) an incoming particle 
with energy $E$ in lead 
$\beta$ and channel $m$ and where the diagonal and
non-diagonal density of states elements
${\cal N}_{\beta\gamma mn}$ are 
\begin{equation} \label{qop1}
{\cal N}_{\beta\gamma mn} (E, E') =  \frac{1}{2\pi i(E-E')}
\sum_{\alpha}  \left[\delta_{mn} \delta_{\alpha\beta}
\delta_{\alpha\gamma} - \sum_{k} s^{\dagger}_{\alpha\beta mk}(E)
s_{\alpha\gamma kn} (E') \right] .
\end{equation}
In particular, in the zero-frequency limit, 
we find \cite{plb} (in matrix notation), 
\begin{equation} \label{qop2}
{\bf {\cal N}}_{\beta\gamma} (E) =  \frac{1}{2\pi i}
\sum_{\alpha}  
{\bf s}^{\dagger}_{\alpha\beta}(E)
\frac{d{\bf s}_{\alpha\gamma} (E)}{dE} ,
\end{equation}
Eq. (\ref{qop2}) is nothing but the multi-lead generalization of 
Eq. (\ref{ws}). 
Note that here we have implicitly assumed that the integration volume 
also enters the scattering matrices used: The integration volume 
intersects the leads of the conductor and we define the scattering 
matrix in such a way that it relates incoming and outgoing waves 
at these intersections. 
Proceeding as for the case of current fluctuations \cite{leso,mb92,physr}
we find for the fluctuation spectrum of the total charge 
\begin{eqnarray} \label{snn}
S_{NN} (\omega) = {h}
\sum_{\gamma\delta} \int dE \,
Tr[{\cal N}^{\dagger}_{\gamma\delta}
(E, E + \hbar\omega) {\cal N}_{\delta\gamma}(E + \hbar\omega, E)]
F_{\gamma \delta} (E, \omega)
\end{eqnarray}
where the trace is over quantum channels and 
\begin{eqnarray} \label{fgd}
F_{\gamma \delta} (E, \omega) = 
\left(f_{\gamma} (E) \left[ 1 - f_{\delta} (E +
\hbar\omega) \right] + \left[ 1 - f_{\gamma} (E) \right] f_{\delta}
(E + \hbar\omega)\right) ,   
\end{eqnarray}
is a combination of frequency-dependent Fermi functions. 
Eq. (\ref{snn}) is, in the absence of 
interactions, the general fluctuation spectrum of the charge 
on a mesoscopic conductor. The true charge on a mesoscopic conductor 
cannot be found from an analysis of free-particle fluctuations. 
As we have pointed out the charge operator and the fluctuation 
spectra depend on the integration volume. Screening is necessary, if only 
to eliminate this dependence. In some particular cases, the volume of 
integration can be physically motivated, for instance in the case of a 
cavity connected to contacts, the main effect we are 
interested in comes from the charging of the cavity itself and thus we
limit the integration volume to that of the cavity. 

Our next goal is to find the true charge, i. e. the charge that 
can build up on a mesoscopic conductor in the presence of interaction. 
Let us consider the low-frequency limit. Then the charge and the potential 
for the conductor shown in Fig. \ref{mesocap} are, on the one hand, related by 
$Q = C U$. Now we write this equation in operator form with a 
charge operator $ \hat Q$ and a potential operator $\hat U$.
We thus have ${\hat Q} = C {\hat U}$. 
On the other hand, the charge on the mesoscopic conductor 
can also be expressed in terms of 
the charge $e \hat{\cal N}$ injected from the reservoir 
(in response of an increase of the 
the reservoir voltage or due to a fluctuation), 
and a screened charge $e^{2} N \hat U$ which is the charge due to the response
of the electrostatic potential to the injected charge. 
Here $N$ is the average density of states. 
Thus in terms of the operators, we have 
\begin{equation} 
{\hat Q} = C {\hat U} = e {\hat {\cal N}} - e^{2} N {\hat U} .
\end{equation} 
This equation now permits us to express $ {\hat U} $ 
in terms of the capacitance and the operator for the bare 
injected charges,  ${\hat U} = e (C+e^{2} N)^{-1} \hat {\cal N}$
and gives  
\begin{equation} 
{\hat Q} = \frac{C e {\hat {\cal N}}}{(C+e^{2} N)} = e \frac{C_{\mu}}{e^{2}}
\frac{{\hat {\cal N}}}{N} 
\label{dip}
\end{equation} 
for the operator of the true charge. Here  
$C_{\mu}$ is the electrochemical capacitance, Eq. (\ref{cmu}).
Eq. (\ref{dip}) states that a charge injected into the conductor 
due to the population of incident states in an energy interval $dE$
is not the bare charge $eNdE$ but only $eNdE/N({e^{2}}/{C_{\mu}})
= {C_{\mu}} dE/e$. 

Next we can use these expression to evaluate 
the (zero-frequency) fluctuation spectrum of the charge. 
We find 
\begin{equation} 
S_{QQ} (0) = \frac{C^{2}_{\mu}}{e^{2}} \frac{S_{NN}(0)}{N^{2}} .
\end{equation} 
Comparison with Eq. (\ref{sqq}) gives us an expression for the
charge relaxation resistance in terms of the fluctuation 
spectrum of the bare charges, 
\begin{equation} 
R_{q} = \frac{1}{2kT e^{2}} \frac{S_{NN}(0)}{N^{2}} .
\label{rqsnn}
\end{equation} 
At equilibrium the particle fluctuation spectrum is 
$S_{NN}(0) = 2 kT h \,$$ Tr({\hat {\cal N}^{\dagger}}{\hat {\cal N}})$
and we recover for $R_q$ the expression, Eq. (\ref{rq}). 
  
As an example let us again consider the QPC as specified by 
the scattering matrix, Eq. (\ref{sqpc}). 
For the elements of the density of states matrix, Eq. (\ref{qop2}), a 
little algebra leads to 
\begin{eqnarray}
    {\cal N}_{11} &=& {\cal N}_{22} = 
    \frac{1}{2\pi} \frac{d\phi_{n}}{dE} ,\\
    {\cal N}_{12} &=& {\cal N}_{21} = \frac{1}{4\pi}  
    \frac{1}{\sqrt{R_{n}T_{n}}} \frac{dT_{n}}{dE} .
\label{mqpc}
\end{eqnarray}
With these density of states elements, we can determine the 
particle fluctuation spectrum, Eq. (\ref{snn})
in the white-noise limit, and  $R_q$ with the help of 
Eq. (\ref{rqsnn}). That leads again to $R_q$ as given by Eq. (\ref{rqgen})
and Eq. (\ref{rqsad}).

\section{The mesoscopic capacitor: mesoscopic gate}

Consider next the case of a system in which all components 
are mesoscopic. An example is the conductor shown 
in Fig. \ref{ped}. 
A mesoscopic cavity is connected (via a single lead) 
to a reservoir and in proximity to a QPC. 
These two conductors can again be viewed as the two plates 
of a small capacitor. 
We assume that each electric field line emanating from 
the cavity ends up either again on the cavity or on the QPC. 
Now the fluctuations of the charge $Q_1$ 
on the cavity and the charge $Q_2$ on the QPC 
are related to the electrostatic potentials $U_1$ and $U_2$ on 
these two conductors via ${Q}_1 = C({U}_{1} - {U}_{2})$
and ${Q}_{2} = C({U}_{2} - {U}_{1})$. 
Note, overall there is no charge accumulation, $Q_{1} + Q_{1}= 0$.
We now present a discussion of the fluctuations 
of the charge fluctuations of this conductor,
generalizing the approach outlined above.

The charge on conductor $i$ can also be written in terms 
of the bare fluctuating charges $e {\hat {\cal N}}_{i}$,
counteracted by a screening charge $e N_{i} ed{\hat U}_{i}$,  
\begin{equation}
{\hat Q}_{1} = C({\hat U}_{1}-{\hat U}_{2}) = 
e {\hat {\cal N}}_{1} - e^{2} N_{1} {\hat U}_{1},
\label{f1}
\end{equation} 
\begin{equation}
{\hat Q}_{2} = C({\hat U}_{2}-{\hat U}_{1}) =
e {\hat {\cal N}}_{2} - e^{2} N_{2} {\hat U}_{2} . 
\label{f2}
\end{equation} 
Note that the charge conservation immediately leads to 
${\hat Q}_{2} = - {\hat Q}_{1}$. We have a dipole and 
$\hat Q$ is the charge operator of the dipole.
We solve these equations for 
${\hat U}_{1}$ and ${\hat U}_{2}$. 
Using $D_{i} \equiv e^{2}N_{i}$ for the density of states
we find that the
effective interaction $G_{ij}$ between the two
systems is
\begin{equation}
    {\bf G} = \frac{C_{\mu}}{D_{1}D_{2}C} \left( \begin{array}{ll}

        C+D_{2} & C \\

        C &  C+D_{1}

        \end{array} \right)
        \label{geff} .
\end{equation}
The electrochemical capacitance is 
the series capacitance of the geometrical contribution and the 
quantum contribution of the two conductors \cite{btp}, 
\begin{equation}
C_{\mu}^{-1} = C^{-1} + (e^{2}N_{1})^{-1}+  (e^{2}N_{2})^{-1} .
\end{equation}
With Eq. (\ref{geff}) we find for the potential operators
\begin{equation}
\hat U_{i} = e \sum_j G_{ij} \hat{ \cal N}_{j} .
\label{uop}
\end{equation}
In the low frequency limit of interest here the elements of the
density of states matrix ${\cal{N}}^{(i)}_{\gamma\delta}$
are specified by Eq. (\ref{qop2}). Using Eqs. (\ref{uop}) and (\ref{snn})
we find that at equilibrium the low frequency fluctuations of the potential
in conductor $1$ are given by
\begin{equation}
S_{U_{1}U_{1}} (0) = 2 kT
\left(\frac{C_{\mu}}{C}\right)^{2}
\left(\left(\frac{C+D_{2}}{D_{2}}\right)^{2} R^{(1)}_{q}
+ \left(\frac{C}{D_{1}}\right)^{2} R^{(2)}_{q} \right) 
\label{su12}
\end{equation}
with $R^{(i)}_{q}$ determined by Eq. (\ref{rq}) using the scattering 
matrix of conductor $i$. Similarly,  
\begin{equation}
S_{U_{2}U_{2}} (0) = 2 kT
\left(\frac{C_{\mu}}{C} \right)^{2}
\left(\left(\frac{C}{D_{2}} \right)^{2} R^{(1)}_{q} + 
\left(\frac{C+D_{1}}{D_{1}}\right)^{2} R^{(2)}_{q} \right) .
\label{su2}
\end{equation}
The voltage fluctuations are correlated, 
\begin{equation}
S_{U_{1}U_{2}} (0) = 2 kT
\left(\frac{C_{\mu}}{C}\right)^{2} 
\left(\frac{(C+D_{2})C}{D_{2}^{2}} R^{(1)}_{q}
+ \frac{(C+D_{1})C}{D_{1}^{2}} R^{(2)}_{q} \right) . 
\label{su122}
\end{equation}
With the help of these spectra we also obtain the 
spectrum $S_{UU}(0)$ of the potential difference $U = U_{1} - U_{2}$, 
which is simply given by 
\begin{equation}
S_{UU} (0) = 2 kT
\left(\frac{C_{\mu}}{C}\right)^{2} 
\left( R^{(1)}_{q}+ R^{(2)}_{q} \right) . 
\label{suu1}
\end{equation}
The charge relaxation resistances of the individual systems add \cite{btp}. 
In the non-interacting limit (infinite capacitance $C$)
the spectra, Eqs. (\ref{su12}-\ref{su122}), all tend to 
the same limit, 
\begin{equation}
S_{U_{i}U_{j}} (0) = 2 kT
\frac{1}{(D_{1}+D_{2})^{2}}
\left( D_{1}^{2} R^{(1)}_{q} + 
D_{2}^{2} R^{(2)}_{q} \right) , 
\label{suinf}
\end{equation}
whereas the spectrum of the voltage difference $S_{UU}$ vanishes.
In the physically more important, charge neutral limit,
we have for the auto-correlation spectra 
\begin{equation} 
S_{U_{i}U_{i}} (0) = 2 kT R^{(i)}_{q}
\label{suinf1}
\end{equation}
and the correlation spectrum 
Eq. (\ref{su122}) vanishes. The fluctuation spectrum of the voltage 
difference is given by Eq. (\ref{suu1}) with $C_{\mu}/C =1$, i. e. 
it is determined just by the sum of the two charge relaxation resistances. 
With the help of these spectra, we can now find the dephasing rates 
in the two conductors. A carrier in conductor $1$ is subject 
to the fluctuating potential $U_{1}$ and a carrier in conductor 
$2$ sees the fluctuating potential $U_{2}$. 
Thus the dephasing rates \cite{mbam} are 
$\Gamma^{(i)}_{\phi} = e^{2}/\hbar^{2} 
\langle \int dt^{\prime} (U_{i}(t^{\prime}) - U_{i}(0))^{2} \rangle$. 
The fluctuation spectra $S_{U_{1}U_{1}}$  and $S_{U_{2}U_{2}}$
each contain two contributions which are additive. For instance, 
the first term in $S_{U_{1}U_{1}}$ represents a contribution 
to the fluctuation spectrum which can be viewed as being predominantly
due to fluctuations in the conductor $1$ itself, whereas the 
second term is due to the presence of the second 
conductor. We can separate the dephasing rate of each conductor 
into two contributions, $\Gamma^{i}_{\phi} = \Gamma^{i1}_{\phi}
+  \Gamma^{i2}_{\phi}$ where the first (upper) index $i$ 
indicates the conductor and the second upper index indicates 
whether this rate is due to the conductor itself (index $ii$)
or due to the presence of the other conductor (index $ij$ with $i \ne j$). 
We thus obtain the following dephasing rates
\begin{eqnarray}
\Gamma^{11}_{\phi} & = & \frac{e^{2}}{h^{2}} kT
\left(\frac{C_{\mu}}{C}\right)^{2}
\left(\frac{C+D_{2}}{D_{2}}\right)^{2} R^{(1)}_{q} , \\
\label{ga11}
\Gamma^{12}_{\phi} & = & \frac{e^{2}}{h^{2}} kT
\left(\frac{C_{\mu}}{D_{1}}\right)^{2} R^{(2)}_{q} , \\
\Gamma^{21}_{\phi}  & = & \frac{e^{2}}{h^{2}} kT
\left(\frac{C_{\mu}}{D_{2}}\right)^{2} R^{(1)}_{q} , \\
\Gamma^{22}_{\phi} & = & \frac{e^{2}}{h^{2}} kT
\left(\frac{C_{\mu}}{C}\right)^{2}
\left(\frac{C+D_{1}}{D_{1}}\right)^{2} R^{(2)}_{q} . 
\end{eqnarray}
In the large capacitance limit, 
this gives for the dephasing rates, 
\begin{equation}
\Gamma^{11}_{\phi}= \frac{e^{2}}{h^{2}} kT
\left(\frac{D_{1}}{D_{1}+D_{2}}\right)^{2} R^{(1)}_{q} , 
\label{ga11i}
\end{equation}
\begin{equation}
\Gamma^{12}_{\phi}= \frac{e^{2}}{h^{2}} kT
\left(\frac{D_{2}}{D_{1}+D_{2}}\right)^{2} R^{(2)}_{q} . 
\label{ga12i}
\end{equation}
In this limit the rates depend on the ratio of the density of states
of the two conductors. In contrast, 
in the zero-capacitance limit, Eq. (\ref{ga11})
gives for the dephasing rate, 
\begin{equation}
\Gamma^{11}_{\phi}= \frac{e^{2}}{h^{2}} kT R^{(1)}_{q}
\label{ga110}
\end{equation}
whereas $\Gamma^{12}_{\phi}$ becomes proportional $C^{2}$
and thus vanishes in the charge-neutral limit. In the important case 
that $C$ is small compared to the quantum capacitance (the Coulomb energy
is large compared to the level spacing)
we find 
\begin{equation}
\Gamma^{12}_{\phi}= \frac{e^{2}}{h^{2}} kT
\left(\frac{C}{D_{1}}\right)^{2} R^{(2)}_{q} .
\label{ga121}
\end{equation}
We have seen that for a single mesoscopic conductor, coupled  to 
a back gate, the treatment of the single lead case, can immediately 
be generalized to the case of a many lead conductor. Similarly, 
the case of two coupled mesoscopic conductors, each of them connected via 
several leads to different reservoirs, is in fact contained in the expressions
given above. We only need to use  
the full scattering matrix of each conductor 
to evaluate the density of states 
and the charge relaxation resistance. This is true only for the 
equilibrium fluctuations. The non-equilibrium situation requires special 
thought.

\section{Nonequilibrium charge fluctuations and dephasing}

\begin{figure}
\epsfxsize=7cm
\centerline{\epsffile{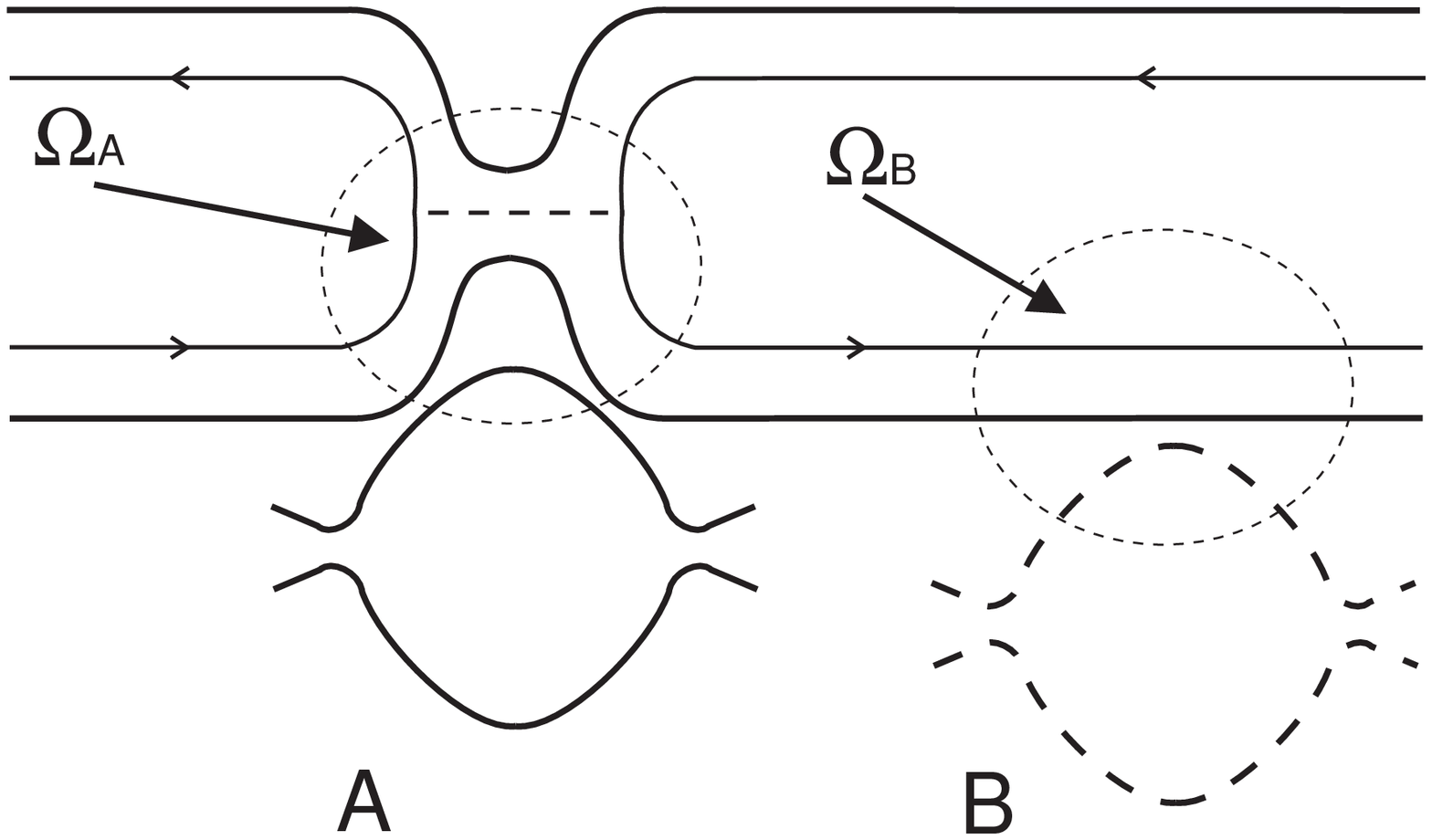}}
\vspace{0.2cm}
\caption{ \label{qpc_geometry}
Quantum point contact coupled to a quantum dot either in position A or B.
After Ref. \protect\cite{mbam}. }
\end{figure}

Let us consider two mesoscopic conductors as above but let 
us consider the case that one of these conductors, say conductor $2$, 
is connected to two reservoirs. 
An example of such a conductor 
is shown in Fig. \ref{qpc_geometry}. One of the conductors, 
a quantum dot or chaotic cavity, is located either in 
position A (to be discussed below) or in position B (to be discussed later
on). A second conductor consists of 
a wire patterned into a two dimensional electron gas. It contains a 
quantum point contact (QPC). The sample is subject to a high magnetic field 
which generates edge states (indicated by lines with arrows). 
In situation A it is the charge on the quantum dot and the charge 
in the center of the QPC which are Coulomb coupled. In situation B 
the charge on the quantum dot is Coulomb coupled with the charge 
on the edge state far away from the QPC. 

Let us now consider the case, where conductor $2$ (the wire 
with the QPC) is brought into a 
transport state. For simplicity, we consider a voltage 
$eV \gg kT $ so large that the thermal noise in this conductor 
can be neglected. To investigate the effect which the presence of 
conductor $2$ has on conductor $1$, we can thus take conductor $2$
to be in the zero-temperature limit. Even
at zero temperature, there is noise generated 
in conductor $2$ which is shot noise due to the granularity of 
the charge \cite{physr}. We are interested in the charge fluctuations 
of this conductor and reconsider Eq. (\ref{snn}). 
In the zero-temperature limit, Eq. (\ref{snn}) is now evaluated 
with a Fermi function for contact $1$ of this conductor 
at an electrochemical potential $\mu_{1}$ 
and a Fermi function for contact $2$ 
at an electrochemical potential $\mu_{2}$. 
To be definite, we take $\mu_{1} > \mu_{2}$ 
such that a voltage $eV = \mu_{1} -\mu_{2}$ is 
established. 
The resulting bare particle number fluctuation spectrum, Eq. (\ref{snn}), 
is now determined only by a subset of the density of states matrix elements
and is proportional to the applied voltage,
\begin{eqnarray} \label{snnv}
S_{NN}^{(2)} (0)  = 2{h} \,
Tr[({\cal N}_{21}^{(2)})^{\dagger} {\cal N}_{21}^{(2)}] e|V| . 
\end{eqnarray}
The upper index $(2)$ is to indicate 
that this spectrum and the matrix elements 
are evaluated for conductor $2$.  
Eq. (\ref{snnv})
is valid to linear order in the applied voltage only. 
To linear order in the applied voltage, we can also in the 
presence of transport use Eqs. (\ref{f1}) and  (\ref{f2}),
since screening is determined by the equilibrium density of states $N$
evaluated at the Fermi energy. The nonequilibrium particle density 
fluctuation spectrum leads to a novel resistance 
\begin{eqnarray}
R^{(2)}_v  = \frac{h}{e^2} \frac{\mbox{Tr} 
    \left( ({\cal N}^{(2)}_{21})^{\dagger} {\cal N}^{(2)}_{21}\right)}
    {N^{2}}.
    \label{Rv} 
\end{eqnarray}
Formally we can define this novel resistance via 
$R_v = e^{2} S_{NN}(0)/2eV$. With this resistance we 
now find for the voltage fluctuation spectra 
\begin{equation}
S_{U_{1}U_{1}} (0) = 2 
\left(\frac{C_{\mu}}{C}\right)^{2}
\left(\left(\frac{C+D_{2}}{D_{2}}\right)^{2} R^{(1)}_{q} kT
+ \left(\frac{C}{D_{1}}\right)^{2} R^{(2)}_{v} e|V| \right) 
\label{su11}
\end{equation}
with $R^{(i)}_{q}$ determined by Eq. (\ref{rq}). For the voltage fluctuation 
spectrum in conductor $2$ we find, 
\begin{equation}
S_{U_{2}U_{2}} (0) = 2 
\left(\frac{C_{\mu}}{C}\right)^{2}
\left(\left(\frac{C}{D_{2}}\right)^{2} R^{(1)}_{q} kT + 
\left(\frac{C+D_{1}}{D_{1}}\right)^{2} R^{(2)}_{v} e|V| \right) .
\label{su21}
\end{equation}
The voltage fluctuations are correlated,  
\begin{equation}
S_{U_{1}U_{2}} (0) = 2 
\left(\frac{C_{\mu}}{C}\right)^{2} 
\left(\frac{(C+D_{2})C}{D_{2}^{2}} R^{(1)}_{q} kT 
+ \frac{(C+D_{1})}{D_{1}^{2}} R^{(2)}_{v} e|V| \right) . 
\label{su121}
\end{equation}
With the help of these spectra we also obtain the 
spectrum $S_{UU}(0)$ of the potential difference $U = U_{1} - U_{2}$, 
which is simply given by 
\begin{equation}
S_{UU} (0) = 2 
\left(\frac{C_{\mu}}{C}\right)^{2} 
\left( R^{(1)}_{q} kT + R^{(2)}_{v} e|V|  \right). 
\label{suu2}
\end{equation}
Whereas the dephasing rate $\Gamma^{11}_{\phi}$ 
remains unchanged and is given by Eq. (\ref{ga11})
the dephasing rate $\Gamma^{12}_{\phi}$ is now 
determined by $R_v eV$ and is given by 
\begin{equation}
\Gamma^{12}_{\phi} = \frac{e^{2}}{\hbar^{2}}
\left(\frac{C_{\mu}}{D_{1}}\right)^{2} R^{(2)}_{v} e|V|
\label{ga12v}
\end{equation}
In the large capacitance limit $({C_{\mu}}/{D_{1}})^{2}$
approaches $({D_{2}}/{D_{1}+ D_{2}})^{2}$ whereas in the small 
capacitance limit $({C_{\mu}}/{D_{1}})^{2}$ tends to $(C/{D_{1}})^{2}$. 

\section{The resistance $R_v$ of a Quantum Point Contact}

We now discuss $R_v$ for the specific example 
of a QPC. The calculation applies to the geometry of Fig. \ref{ped}
or to the geometry of Fig. \ref{qpc_geometry} 
with the quantum dot in position A. 
Using the density matrix elements 
for a symmetric QPC given by Eqs. (\ref{mqpc}),  
we find \cite{plb} 
\begin{eqnarray}
R_v =
\frac{h}{e^2} 
\frac{ \sum_n \frac{1}{4R_{n}T_{n}}
	\left( \frac{dT_{n}}{dE} \right)^{2}}
{[\sum_{n} (d\phi_{n}/dE)]^{2}}	
= {\frac{h}{e^{2}}} 
(\frac{\pi}{\hbar \omega_x})^{2}
\frac{\sum_n T_{n}R_{n}}{[\sum_{n} (d\phi_{n}/dE)]^{2}} ,
\label{rvqpc1}
\end{eqnarray}	
where we have made use of the fact that the transmission probabilities 
have the form of Fermi functions, see Eq. (\ref{dtde}). 
Note that the resistance $R_{v}$ 
is proportional to the shot noise power 
$S(0) = 2(e^{2}/h) \sum_{n} T_{n}R_{n} eV$.

This statement is in contrast to the 
experimental papers \cite{buks1,buks2} and 
to theoretical works \cite{aleiner1} 
where arguments are 
presented 
which lead to a dephasing rate 
proportional to $(\Delta T)^{2}$
$/4TR$. 
The arguments which are advanced take the factor 
$TR$ as being due to the shot noise 
and $\Delta T$ as the change in the transmission probability 
due to the variation of the charge on the QPC. According to these
arguments, the dephasing rate is inversely proportional to the shot noise 
power $S$. Of course we are not forbidden to relate $TR$ 
to the zero temperature 
current noise spectrum. But this identification
breaks down, if we are not strictly at zero
temperature, if the channels of the QPC 
do not open in a well separated way 
(and several channels contribute) etc. 
More importantly, as the quantum dot 
is moved from position A in Fig. \ref{qpc_geometry} 
to position B the resulting dephasing rate is, as we will see, 
unambiguously proportional to the shot noise. Clearly arguments which 
hold that the dephasing rate is inversely proportional to the shot 
noise in configuration A but proportional to the shot noise
in configuration B lead to a paradox. 

Also Eq. (\ref{rvqpc1}) is valid for a 
saddle point constriction, the adiabatic model of Glazman et al. \cite{glaz}
leads to the same conclusion. Thus it stands to reason that 
Eq. (\ref{dtde}) is in fact much more general than the simple model used 
indicates. 

\begin{figure}
\epsfxsize=7.5cm
\centerline{\epsffile{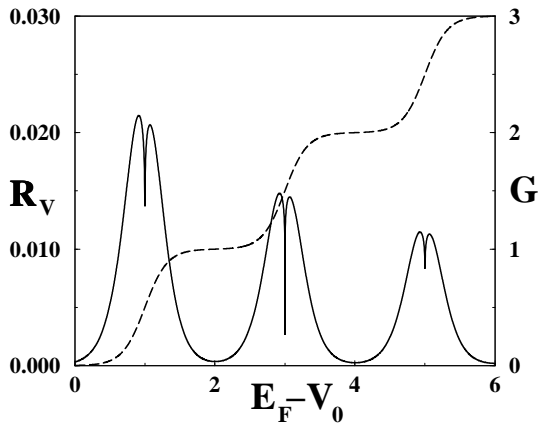}}
\vspace{-0.5cm}
\caption{ \label{rvedge}
$R_{v}$ (solid line) for a saddle QPC 
in units of ${h/e^{2}}$ and $G$ (dashed line)
in units of ${e^{2}/h}$ as a function of $E_{F} - V_{0}$ in units of 
$\hbar \omega_x$ 
with $\omega_y/\omega_x = 2$ 
and a screening length $m \omega_x \lambda^{2}/\hbar = 25$.
$R_v$ and $G$ are for spinless electrons. 
After Ref. \protect\cite{ammb}. 
}
\end{figure}

To evaluate Eq. (\ref{rvqpc1}) further,  
Ref. \cite{plb} calculates 
the phase derivatives $d\phi_{n}/dE$ semi-classically 
in WKB approximation. The resulting resistance $R_v$ is shown in 
Fig. \ref{rvedge}. Application of a magnetic field changes 
$R_v$ only qualitatively (see Ref. \cite{mbam}).  

Without screening $R_v$ would exhibit a bell shaped behavior as a function 
of energy, i. e. it would be proportional to $T_{n}(1-T_{n})$
in the energy range in which the n-th transmission channel is 
partially open. Screening, which in $R_v$ is inversely proportional to the 
density of states squared, generates the dip at the threshold 
of the new quantum channel at the energy which corresponds to $T_{n} = 1/2$. 
In the semi-classical evaluation 
used here the density of states is singular at this point. 
Quantum corrections will reduce the density of states 
and will tend to diminish the dip shown in Fig. \ref{rvedge}. 
It is interesting to note that the experiment \cite{buks1} does 
indeed show a double hump behavior of the dephasing rate. 

\section{Local Charge Fluctuations} 

Thus far we discussed situations in which the total charge 
fluctuation determine the potential fluctuations and the dephasing 
rates. 
Many additional geometries, however, require a discussion not only 
of the total charge fluctuations, but a consideration of the local charge.

To describe the charge distribution due to carriers 
in an energy interval $dE$ in our conductor, we consider the 
Fermi-field \cite{mb92}
\begin{equation}\label{fermi}
\hat{\Psi} ({\bf r},t)  = \sum_{\alpha m} \int dE 
(1/hv_{\alpha m} (E))^{1/2} 
\psi_{\alpha m} ({\bf r},E)
\hat{a}_{\alpha m} (E) exp(-iEt/\hbar), 
\end{equation}
which annihilates an electron at point ${\bf r}$ and time $t$.
The Fermi field Eq. (\ref{fermi}) is built up 
from all scattering states $\psi_{\alpha m} ({\bf r},E)$
which have unit incident amplitude in contact $\alpha$ in channel
$m$. The operator $\hat{a}_{\alpha m} (E)$ annihilates 
an incident carrier in reservoir $\alpha$ in channel
$m$. $v_{\alpha m}$ is the velocity in the incident channel $m$ in 
reservoir $\alpha$. 
The local carrier density at point ${\bf r}$ and time $t$
is determined by $\hat{n}({\bf r},t) = \hat{\Psi}^{\dagger}({\bf r},t)
\hat{\Psi}({\bf r},t)$. 
We will investigate 
the density operator in the frequency domain, $\hat{n}({\bf r},\omega)$.
Using the Fermi-field we find, 
\begin{eqnarray}\label{denp1}
\hat{n}({\bf r},\omega) =
& &\sum_{\alpha m \beta n} \int dE 
(1/hv_{\alpha m} (E))^{1/2}
(1/hv_{\beta n} (E + \hbar \omega))^{1/2} \nonumber \\
& & \psi_{\alpha m}^{\ast} ({\bf r},E)  
\psi_{\beta n} ({\bf r},E + \hbar \omega) 
\hat{a}_{\alpha m}^{\dagger} (E) 
\hat{a}_{\beta n}(E + \hbar \omega) .
\end{eqnarray}
This equation defines a density matrix operator with elements 
\begin{eqnarray}\label{denp2}
{n}_{\gamma \delta  mn}({\bf r}, E, E + \hbar \omega) = && h^{-1}
(v_{\gamma m} (E) v_{\delta n} (E + \hbar \omega))^{-1/2} \nonumber \\
&& \psi_{\gamma m}^{\ast} ({\bf r},E) 
\psi_{\delta  n} ({\bf r},E + \hbar \omega) .
\end{eqnarray} 
It is now very convenient and instructive to consider an expression 
for the density operator not in terms of wave functions 
but more directly in terms of the scattering matrix. 
In the zero-frequency limit, 
in matrix form, we can connect the scattering states and 
the scattering matrix with the density of states matrix \cite{math}, 
\begin{equation} 
{\bf n}_{\beta \gamma} (\alpha, {\bf r}) = 
- (1/4\pi i)
[{\bf s}^{\dagger}_{\alpha\beta}
(\delta {\bf s}_{\alpha \gamma}/\delta eU({\bf r})) -
(\delta {\bf s}^{\dagger}_{\alpha \beta}/\delta eU({\bf r})) 
{\bf s}_{\alpha \gamma}] .
\label{elmdef}
\end{equation} 
All quantities in this expression are evaluated at the energy $E$.
The matrix elements of Eq. (\ref{denp2}) 
are connected to the matrices ${\bf n}_{\beta \gamma} (\alpha, {\bf r})$ 
via 
\begin{equation}
{\bf n}_{\gamma \delta}({\bf r}) = \sum_{\alpha}
{\bf n}_{\beta \gamma} (\alpha, {\bf r})
\label{connect}
\end{equation}
The price we have to pay to gain local information is to use 
functional derivatives of the scattering matrix with 
respect to the local potential. 

Eq. (\ref{connect}) was given in Ref. \cite{math} without proof.
Appendix B outlines how this result is obtained. 
We are  not interested in the microscopic local 
charge fluctuations but we will consider the fluctuating charge 
in a certain volume $\Omega$. To make the problem treatable, 
we will make the assumption that in the volume of interest 
the potential can be described by a single variable. 
Thus the quantities of interest are obtained 
by integrating the density operator Eq. (\ref{connect}) 
over a volume $\Omega$,
\begin{equation}\label{caln}
{\bf {\cal N}}_{\beta \gamma} = 
\int_{\Omega} d^{3}{\bf r} \, {\bf n}_{\beta \gamma} ({\bf r}) . 
\end{equation} 
With the help of 
the charge density matrix the low frequency limit of the 
bare charge fluctuations can be obtained~\cite{btp,plb,mbam}. It is 
again given by Eq. (\ref{snn}) but now with the local 
density of states matrix:   
the elements of ${\cal N}_{\gamma\delta}$ are in the zero-frequency limit 
of interest here given by Eq. (\ref{caln}). Thus we are now in a position 
to find the local charge fluctuations in any volume element of interest.

\section{Charge Fluctuations of an Edge State} 

Consider the conductor shown in Fig. \ref{qpc_geometry} but 
with conductor $1$ located at B. Now the charge fluctuations
on the quantum dot 
couple to the charge fluctuations on the edge state. 
To simplify the problem, we assume that the relevant
charge fluctuations in conductor $2$ are of importance only 
near the quantum dot (in the region $\Omega_B$ of Fig. \ref{qpc_geometry}).
The rest of the conductor is treated as charge neutral. 

The scattering matrix of the QPC alone can be described by 
$r \equiv s_{11} = s_{22} = - i {\cal R}^{1/2}$
and $t \equiv s_{21} = s_{12} = {\cal T}^{1/2}$
where ${\cal T} = 1- {\cal R}$ is the transmission probability 
through the QPC. 
Here the indices $1$ and $2$ label the reservoirs 
(see Fig. \ref{qpc_geometry}).
We can entirely neglect the phase accumulated 
by carriers traversing the QPC. However, 
a carrier traversing the region underneath the gate 
acquires a phase $\phi(U)$ which depends on the 
electrostatic potential $U$ in this region. 
Since we consider only the charge pile up in this region 
all additional phases
in the scattering problem are here without importance. 
The total scattering
matrix of the QPC and the traversal of the region $\Omega$ is then simply
\begin{equation} \label{sm}
{\bf s} = 
\left( \matrix{ r & t \cr 
t e^{i\phi} & r e^{i\phi} } \right).
\end{equation} 
If the polarity of the magnetic field is reversed 
the scattering matrix is given by $s_{\alpha\beta}(B) = s_{\beta\alpha}(-B) $, 
i. e. in the reversed magnetic field it is only the second column
of the scattering matrix which contains the phase $\phi(U)$. 
In what follows, the dependence of the scattering matrix on the phase $\phi$
is crucial. We emphasize that the approach presented here can be generalized 
by considering all the phases of the problem and by 
considering these phases and the amplitudes to depend 
on the entire electrostatic potential landscape \cite{math}.  
In the evaluation of the density of states matrix elements , 
we make use of the fact that 
$d\phi/edU = - d\phi/dE $ in the WKB-limit.
However, $d\phi/dE = 
2\pi N$ where $N$ is just the density of states of the edge state 
underneath the gate (in region $\Omega_B$  of Fig. \ref{qpc_geometry}).
Simple algebra
gives \cite{mbam,ammb} ${\cal N}_{11} =  {\cal T} N$,
\begin{equation}
{\cal N}_{21} = {\cal N}^{\ast}_{12} = r^{\ast}t N,
\label{n12}
\end{equation}
and ${\cal N}_{22} = {\cal R} N_{2}$.
Using only the zero-frequency 
limit of the elements of the charge operator determined above gives, 
\begin{eqnarray}\label{qfluct2}
S_{NN}(\omega) =  h N^{2} & & \Big[{\cal T}^{2} \int dE\ F_{11}(E,\omega) 
                  + {\cal T}{\cal R} \int dE\ F_{12}(E,\omega)\nonumber\\
                         & & + {\cal T}{\cal R} \int dE\ F_{21}(E,\omega)
                           + {\cal R}^{2} \int dE\ F_{22}(E,\omega) \Big] ,
\end{eqnarray}
which in the zero-frequency limit is, 
\begin{eqnarray}\label{qfluct3}
S_{NN}(\omega) =  2 h N^{2} & & \Big[{\cal T}\int dE\ f_{1}(1-f_{1})
                  + {\cal T}{\cal R} \int dE\ (f_{1} - f_{2})^{2}\nonumber\\
                 & & + \int dE\ {\cal R} f_{2}(1-f_{2}) \Big] .
\end{eqnarray}
The first term represents the equilibrium noise which is transmitted 
from contact $1$ through the QPC into the edge channel adjacent to the 
quantum dot. The last term is the equilibrium noise of reservoir $2$ 
which is fed into this edge state through reflection at the QPC.
The middle term is the non-equilibrium, (two particle) shot noise 
contribution. In the zero temperature limit, the equilibrium 
noise terms do not contribute, and the shot noise term is 
proportional to the applied voltage and given by 
\begin{eqnarray}
    S_{NN}(0) = 2h N^{2} {\cal R}{\cal T}  e|V| .
\label{qfluct4}
\end{eqnarray}
To proceed, we can again consider Eqs. (\ref{f1}) and (\ref{f2}).
The charge operator $d{\hat Q}_{1}$ describes as before the 
charge fluctuations on the quantum dot, $d{\hat Q}_{2}$ the 
charge fluctuations on the edge state in proximity 
to the quantum dot. Now $\hat{\cal N}_{2}$ is the bare 
particle fluctuation operator on the edge state, and $N_{2}$
and $\hat{\cal U}_{2}$ are the density of states of the edge state 
in $\Omega_B$ and the potential operator in the region $\Omega_B$. 
The potential fluctuations in conductor $1$ are again 
given by Eq. (\ref{su11})
but with a resistance  
$R^{(2)}_{v}$ which is \cite{mbam}
\begin{eqnarray}
      R^{(2)}_{v} = (h/e^{2}) {\cal R}{\cal T}.
\label{rqv}
\end{eqnarray}
For a single edge state the resistance $R_v$ is independent 
of the density of states. The resistance $R_v$ is again 
proportional to the shot noise generated by the QPC. 
It is maximal for a semi-transparent QPC, ${\cal T} = {\cal R} =1/2$.  
The resulting dephasing rate is \cite{mbam}
\begin{equation}
\Gamma^{12}_{\phi} = \frac{e^{2}}{\hbar^{2}}
\left(\frac{C_{\mu}}{D_{1}}\right)^{2} R^{(2)}_{v} eV = 4 \pi^{2} 
\left(\frac{C_{\mu}}{D_{1}}\right)^{2}{\cal R} {\cal T}e|V|. 
\label{ga12ve1}
\end{equation} 
Suppose now that the quantum dot is at resonance. 
Then its density of states is given by ${D_{1}} = 2e^{2}/(\pi \Gamma)$
where $\Gamma$ denotes the width of the quantum level due to 
decay of carriers into the leads. 
We then obtain
\begin{equation}
\Gamma^{12}_{\phi} = \pi^{4} \Gamma^{2} 
\left(\frac{C_{\mu}}{e^{2}}\right)^{2}{\cal R}{\cal T}e|V|. 
\label{ga12ve2}
\end{equation} 
In the limit $e^{2}/C>>
\pi \Gamma/2$ and $e^{2}/C \gg N_{2}^{-1}$
the capacitance $C_{\mu}$ can be replaced by the geometrical capacitance 
$C$. In the opposite limit, if   $e^{2}/C <<
\pi \Gamma/2$ and $e^{2}/C << N_{2}^{-1}$ we find 
a dephasing rate 
\begin{equation}
\Gamma^{12}_{\phi} = \pi^{4} (N_{2}\Gamma)^{2}
{\cal R}{\cal T}e|V|. 
\label{ga12ve3}
\end{equation} 
The limit described by 
Eq. (\ref{ga12ve2}) is realistic for mesoscopic samples. 

The dephasing rate given by Eq. (\ref{ga12ve2}) is inversely proportional to
the square of the coupling constant $e^{2}/C$. It vanishes as this 
coupling constant tends to infinity, since charge fluctuations 
become energetically prohibitive. In contrast, a perturbation 
treatment which considers only the bare charge fluctuations 
would lead to a result that is proportional to the square 
of the coupling constant. Second, the rate given by Eq. (\ref{ga12ve2})
is proportional to the square of the width of the resonance in the 
quantum dot. 

Ref. \cite{mbam} presents two additional results. 
First if there are more than two edge states which contribute to transport 
the charge resistance $R_v$ is modified. As an example, consider the case, 
where one of the edge states is transmitted perfectly through the conductor. 
Transport in this edge state (in the zero-temperature
limit) generates no current noise. But it can contribute to screening 
if the two edge states are not to far apart near the gate. If both edge 
states see the same potential $U_2$, the resistance $R_v$ becomes, 
\begin{equation}
R_v = \frac{h}{e^{2}} 
\left(\frac{N_{2}}{N_{1}+ N_{2}}\right)^{2} {\cal T}{\cal R}e|V| ,
\label{last}
\end{equation}
where $N_{1}$ is the density of states of the outer edge state 
in $\Omega_B$ and $N_{2}$ is the density of states of the 
the inner only partially transmitted edge state in region 
$\Omega_B$.  
Eq. (\ref{last}) is valid if there is no population equilibration
(due to elastic or inelastic scattering) 
among the two edge channels between the QPC and the dot.
Thus in the presence of additional edge states, the resistance 
$R_v$ is reduced below that of a single edge state. 

Another result given in Ref. \cite{mbam} is of experimental 
interest \cite{buks2}. If a voltage probe is inserted between 
the QPC and the quantum dot in position B, and if the voltage 
probe is such that every carrier enters it, it acts as 
a complete phase randomizer \cite{bu85}. Ref. \cite{mbam} 
shows that for a single edge state $R_v$ and thus the 
dephasing rate, remains unchanged by the presence of the voltage probe, 
if the voltage probe is ideal. An ideal voltmeter has infinite impedance.
To derive this result, it is necessary to know the ac-conductance 
matrix of the conductor, since the voltage at the probe is now 
a time-dependent fluctuating quantity. Its Fourier amplitude acts 
thus like an ac-voltage applied at this contact.
Ref. \cite{mbam}
only stated the result, for a more detailed 
discussion we must refer the reader to Ref. \cite{sitges}.

\section{Discussion}

In this work we have discussed the relationship 
between the dephasing rate 
and quantities which determine the $RC$-time in Coulomb coupled structures. 
The $RC$-time reflects a collective behavior of the electrons in these
two conductors. In fact, as we have pointed out, as far as the total 
charge is concerned, the Coulomb interaction, imposes, via the requirement 
of charge neutrality over large distances, a complete correlation
between the charge fluctuations on the two conductors. The $RC$-time 
is determined by an electrochemical capacitance and at equilibrium 
by a charge relaxation resistance $R_q$ and in the presence of transport 
by a resistance $R_v$. 

We have emphasized the simple case, where each conductor 
is only described by one potential. There is then only 
one dipole whose fluctuations govern the charge dynamics 
of the coupled conductors. The theory is not limited 
to such a simplified discussion. A formal expression for
the electrochemical capacitance \cite{mb93} and for the equilibrium charge 
relaxation resistance \cite{math} using the microscopic potential 
landscape have already been derived. However, for the geometries
which do not allow for any geometrical symmetries, such general 
expressions are difficult to evaluate. More fruitful is an approach, 
which proceeds like the discussion of ac-conductance, by dividing 
the sample into a limited series of volumes whose potentials 
and charges are related by a geometrical capacitance matrix. 
In such a discrete potential model it is then possible with some 
finite algebraic effort, to present a theory which contains not only
the dynamics of one fluctuating dipole, but that of a number of 
dipoles. 

We conclude by mentioning that recently progress has also been made 
to extend the theory presented here to normal-superconducting, hybrid 
structures \cite{tgmb,amtgmb}. 
This is interesting since charge and particle fluctuations 
due to the presence of electron and hole quasi-particles are not simply
proportional to one another as in a normal conductor. 

\section*{Acknowledgement}

I have benefited from discussions with B. L. Altshuler, Y. M. Blanter, 
C. W. J. Beenakker, A. M. Martin and C. Texier. 
This work is supported by the Swiss
National Foundation and the European Network on Dynamics of 
Hybrid Nanostructures.  

\section*{Appendix A: The current operator}

In this appendix we recall some results form the scattering 
theory of electrical conduction which are needed in the main part 
of this work. We consider a conductor and assume 
that its internal electrostatic potential is fixed at its static 
equilibrium value. The reservoirs are represented as perfect 
conductors which permit a sequence of transverse states
with threshold energy below the Fermi energy. These states form 
the quantum channels which permit the definition of in-going and outgoing 
particle fluxes. The current operator in such a reservoir is, 
\begin{equation}
\hat{I}_\alpha (\omega) = {e} \int dE [
\hat{a}^{\dagger}_{\alpha} (E) \hat{a}_{\alpha} (E+\hbar\omega) -
\hat{b}^{\dagger}_{\alpha} (E) \hat{b}_{\alpha} (E+\hbar\omega) ]
\label{cur1}
\end{equation}
where $\hat{a}^{\dagger}_{\alpha}(E)$ creates an incoming  particle flux 
in reservoir $\alpha$. $\hat{a}_{\alpha}(E)$ is an $M_{\alpha}$
component vector: one component for each transverse channel 
with threshold below the Fermi energy. $\hat{b}^{\dagger}_{\alpha} (E)$
creates out-going fluxes in lead $\alpha$. Eq. (\ref{cur1})
is just the Fourier transform in frequency-space of the 
time-dependent occupation number 
of the incoming currents minus the occupation number of the 
outgoing currents. The operators $\hat{a}_{\alpha}(E)$ 
and $\hat{b}_{\alpha} (E)$ are not independent but related 
by a unitary transformation \cite{mb92}:
\begin{equation}
\hat{b}_\alpha = \sum_\beta {\bf s}_{\alpha\beta} \hat{a}_\beta .
\label{smatrix}    
\end{equation}
Here  ${\bf s}_{\alpha\beta}$ is 
the scattering matrix 
with dimensions $M_{\alpha}\times M_{\beta}$
which relates the incoming (current) amplitudes to the outgoing
current amplitudes. 
Using Eq.\ (\ref{smatrix}) to eliminate the occupation numbers of the 
outgoing channels in terms of the ingoing channels 
yields a current operator \cite{mb92} 
\begin{equation}
\hat{I}_{\alpha}(\omega) = {e} \int dE \sum_{\beta\gamma}
\hat{a}_\beta^\dagger(E) {\bf A}_{\beta\gamma}(\alpha,E,E+\hbar\omega)
\hat{a}_\gamma(E+\hbar\omega).
\label{curo}
\end{equation}
with a {\em current matrix}  
\begin{equation}
{\bf A}_{\delta\gamma}(\alpha,E,E^{\prime}) = \delta_{\alpha\delta}
\delta_{\alpha\gamma} {\bf 1}_\alpha - 
{\bf s}_{\alpha\delta}^\dagger(E) {\bf s}_{\alpha\gamma}(E^{\prime}).
\label{curm}
\end{equation}
Here ${\bf 1}_\alpha$ is the unit matrix with dimensions $M_\alpha \times
M_\alpha$. 
The properties of the current matrix are discussed in detail in
Ref. \cite{mb92}.

\section*{Appendix B: Scattering matrix expression for local densities} 

To derive Eq. (\ref{connect}), we start 
from the Schr\"odinger equation for carriers 
in an electrostatic potential $U({\bf r})$.
The scattering state $\psi_{\delta}({\bf r},E)$
is an exact solution of this equation. 
Now we add a small complex valued potential $ - i \Gamma ({\bf r})$ to the real 
electrostatic potential $U({\bf r})$. The complex valued potential 
is non-vanishing only in a small region around ${\bf r}$. 
The complex valued potential changes the scattering state 
$\psi_{\delta}({\bf r},E)$ due to absorption at ${\bf r}$ into a state 
$\Psi_{\delta}({\bf r},E, \Gamma ({\bf r}))$. Now we proceed by multiplying
the Schr\"{o}dinger equation for 
$\Psi_{\delta} ({\bf r},E, \Gamma ({\bf r}))$ by  
$\Psi_{\gamma}^{\ast}({\bf r},E, \Gamma ({\bf r}))$.
Next we consider the Schr\"{o}dinger equation for the state 
$\Psi_{\gamma}^{\ast}({\bf r},E, \Gamma ({\bf r}))$. 
The potential has a small positive 
imaginary part at ${\bf r}$, and thus a total potential 
$eU({\bf r}) + i \Gamma ({\bf r})$. 
We multiply this equation with $\Psi_{\delta}({\bf r},E, \Gamma ({\bf r}))$.
Subtraction of the two equations gained in this manner from one another
gives
\begin{equation} 
- (\hbar^{2} /2m) 
[\Psi^{\ast}_{\gamma} \Delta^{2} \Psi_{\delta}-
\Psi_{\delta} \Delta^{2} \Psi^{\ast}_{\gamma}] =
-2i \Gamma ({\bf r}) \Psi^{\ast}_{\gamma} \Psi_{\delta}  . 
\end{equation}
The left hand side of this equation is, apart from a factor
$i\hbar$ just the divergence of the local current density
$\tilde{j}_{\gamma\delta}({\bf r})$. 
Thus we can also write 
\begin{equation} 
\hbar \nabla \tilde{j}_{\gamma\delta}({\bf r}) =
-2\Gamma ({\bf r})
\Psi^{\ast}_{\gamma}\Psi_{\delta} .
\end{equation}
The current-matrix elements which are related to the scattering matrix
correspond to a carrier flux not at some definite energy but 
to the current in some small energy interval, 
$j_{\gamma\delta}(E^{\prime},E,{\bf r})  = 
(1/hv_{\gamma} (E^{\prime}))^{1/2}
(1/hv_{\delta} (E) )^{1/2}$
$\tilde{j}_{\gamma\delta}(E^{\prime},E, {\bf r}) dE^{\prime} dE$.
Using this (for $E = E^{\prime}$) 
and integrating the resulting equation over the volume of the 
conductor, gives
\begin{equation}
\sum_{\alpha} d{\bf I}_{\gamma\delta}(\alpha) =   
2 \int d^{3}r (1/hv_{\gamma})^{1/2}
(1/hv_{\delta})^{1/2} \Gamma ({\bf r}) 
\Psi^{\ast}_{\gamma} \Psi_{\delta} .
\end{equation}
where we have taken currents which enter the volume to be positive. 
Here we have used the freedom in the integration volume to make it 
large enough such that the surface of the volume intersects the reservoirs. 
If these intersections coincide with the intersections which are used
to define the scattering matrices, we can use 
the expressions which relate the currents to the scattering matrix elements
\begin{equation}
\sum_{\alpha} {\bf A}_{\gamma\delta}(\alpha, \Gamma) = 2  
\int d^{3}r (1/hv_{\gamma})^{1/2}
(1/hv_{\delta})^{1/2} \Gamma ({\bf r}) 
\Psi^{\ast}_{\gamma} \Psi_{\delta} .
\end{equation}
Now we take on both sides the functional derivative with respect to
$\Gamma$. At $\Gamma = 0$, on the right hand side,
this derivative gives us back 
the original scattering 
states which exist in the absence of absorption, 
\begin{equation}
(1/h)\sum_{\alpha} \delta {\bf A}_{\gamma\delta}(\alpha)
/\delta\Gamma({\bf r}) |_{\Gamma = 0} =
(4\pi/h^{2}) 
(v_{\gamma}v_{\delta})^{-1/2} 
\psi^{\ast}_{\gamma}\psi_{\delta} .
\end{equation}
It remains to re-write the functional derivative 
$\delta {\bf A}_{\gamma\delta}(\alpha)
/\delta\Gamma({\bf r})$ in terms of functional derivatives 
with respect to the potential. To do this we note that 
in the presence of absorption the scattering matrix ${\bf s}$ 
is a functional of 
$eU({\bf r}) -i\Gamma ({\bf r})$  and the scattering matrices 
${\bf s}^{\dagger}$ 
are a functional of $eU({\bf r}) + i\Gamma ({\bf r})$. It follows that 

\begin{equation}
\delta {\bf A}_{\gamma\delta}(\alpha)/\delta\Gamma({\bf r}) = 
i (\delta {\bf s}^{\ast}_{\alpha\gamma}/e\delta 
U({\bf r})) {\bf s}_{\alpha\delta} 
- {\bf s}^{\ast}_{\alpha\gamma}(\delta {\bf s}_{\alpha\delta}/\delta 
U({\bf r}))
\end{equation}
and hence 
\begin{equation}
\sum_{\alpha} n_{\gamma\delta}(\alpha ,{\bf r}) = (1/h)
(v_{\gamma}v_{\delta})^{-1/2} 
\psi^{\ast}_{\gamma}\psi_{\delta} .
\label{den2}
\end{equation}

\end{document}